\documentclass[twocolumn,superscriptaddress,prl]{revtex4}
\usepackage{mathrsfs, doi, hyperref, dsfont,mathtools}
\usepackage{amssymb, amsbsy, amsmath, latexsym, mathrsfs,dsfont, array, layout, graphicx,mathrsfs,braket,amsfonts,amsthm, amssymb,graphicx,subfigure, youngtab,color,xcolor,bm}
\usepackage{times}

\def\>{\rangle}
\def\<{\langle}

\newcommand{\bi}[2]{\binom{#1}{#2}}

\newcommand{\bsym}[1]{\boldsymbol{#1}}

\hypersetup{
    colorlinks=true,  
    linkcolor=blue, 
    citecolor=blue, 
    filecolor=blue,
    urlcolor=blue
}

\begin{document}

\title{A quantum approach to  homomorphic encryption}

\author{Si-Hui Tan}
\affiliation{Singapore University of Technology and Design, 8 Somapah Road, Singapore 487372}
\author{Joshua A. Kettlewell}
\affiliation{Singapore University of Technology and Design, 8 Somapah Road, Singapore 487372}
\author{Yingkai Ouyang}
\affiliation{Singapore University of Technology and Design, 8 Somapah Road, Singapore 487372}

\author{Lin Chen}
\affiliation{School of Mathematics and Systems Science, Beihang
University, Beijing 100191, China}
\affiliation{International Research Institute for Multidisciplinary
Science, Beihang University, Beijing 100191, China}
\affiliation{Singapore University of Technology and Design, 8 Somapah Road, Singapore 487372}

\author{Joseph F. Fitzsimons}
\email{joseph\_fitzsimons@sutd.edu.sg}

\affiliation{Singapore University of Technology and Design, 8 Somapah Road, Singapore 487372}
\affiliation{Centre for Quantum Technologies, National University of Singapore, 3 Science Drive 2, Singapore 117543}

\begin{abstract}
Encryption schemes often derive their power from the properties of the underlying algebra on the symbols used. Inspired by group theoretic tools, we use the centralizer of a subgroup of operations to present a private-key quantum homomorphic encryption scheme that enables a broad class of quantum computation on encrypted data. 
A particular instance of our encoding hides up to a constant fraction of the information encrypted. This fraction can be made arbitrarily close to unity with overhead scaling only polynomially in the message length. This highlights the potential of our protocol to hide a non-trivial amount of information, and is suggestive of a large class of encodings that might yield better security.
\end{abstract}

\date{\today}
\maketitle

The discovery that quantum systems could be harnessed to process data in a fundamentally new way has led to the burgeoning field of quantum information processing. This approach to computation holds the promise of more efficient algorithms for a variety of tasks including integer factorization \cite{shor1997polynomial}, search  \cite{Grover:1996:FQM:237814.237866} and quantum simulation \cite{lloyd1996universal}. However, quantum information processing has also found applications in the area of cryptography, which has been a focus of the field since the discovery of secure quantum key distribution protocols by Bennett and Brassard \cite{BB84}, and Ekert \cite{PhysRevLett.67.661}. The information theoretic security of these protocols stands in stark contrast to the reliance of classical key agreement protocols on assumptions of computational hardness, and indeed a major goal of quantum cryptography research is to replicate and extend the functionality present in existing classical schemes while providing stronger, information theoretic, security guarantees.

In the world of classical cryptography, a central topic in recent years has been the study of homomorphic encryption \cite{Rivest1978,Gentry:2009:FHE:1536414.1536440,DGH2010}. Homomorphic encryption is a form of encryption which allows data processing to be performed on encrypted data without access to the encryption key. In general, a homomorphic encryption system is composed of four components:  a {\it key generation algorithm}, an {\it encryption algorithm} that encrypts the data using the generated key, a {\it decryption algorithm} that decrypts the data using the key, and an {\it evaluation algorithm} which is used to process the data without decryption. Thus homomorphic encryption allows for secret data to be processed by third parties without allowing them access to the plaintext. After decryption, the plaintext output reveals the processed data. A scheme is termed \textit{fully-homomorphic} if it allows for arbitrary processing of the encrypted data. Although the idea for homomorphic encryption has existed for some time \cite{Rivest1978}, it was not until 2009 that a fully-homomorphic encryption scheme was discovered by Gentry \cite{Gentry:2009:FHE:1536414.1536440}. Gentry's scheme is only computationally secure, relying on the assumed hardness of certain worst-case problems over ideal lattices, and the sparse subset sum problem, although the condition requiring ideal lattices was later dropped \cite{DGH2010}. 

Recent successes in quantum cryptography in finding information theoretically secure protocols for blind computation \cite{5438603, ABE08, Barz20012012, PhysRevA.87.050301, PhysRevLett.111.230501, mantri2013optimal} and verifiable computing \cite{FK13,RUV13,SFKW13,M10}, problems closely linked to homomorphic encryption, have motivated the question of whether quantum mechanics allows for information theoretically secure homomorphic encryption schemes. Indeed, a number of attempts have been made to find a quantum analogue of homomorphic encryption \cite{Liang2013, Liang2014, FBS2014, Childs:2005:SAQ:2011670.2011674}, however these attempts have inevitably run into a barrier. It is now known that it is not possible to achieve perfect information theoretic security while enabling arbitrary processing of encrypted data, unless the size of the encoding is allowed to grow exponentially \cite{YPF2014}. As a result, such schemes have required interaction between parties to enable deterministic computation. These requirements parallel those of blind quantum computation which hides {\it both} the data and the computation being done on it. The question then remains as to whether information theoretically secure homomorphic encryption is possible without expanding the definition to include interactive protocols. A first step in the direction of non-interactive quantum protocols was presented in \cite{PhysRevLett.109.150501} for a restricted model of quantum computation known as the BosonSampling model \cite{AA11} which is non-universal. Furthermore, the scheme ensures only that the encoded information and the accessible information differ by an amount proportional to $\log_2 m$ bits when $m$ bits are encrypted, which is a relatively weak security guarantee. An information-theoretically secure scheme that allows for processing of encrypted data beyond BosonSampling is not known to date.

In this paper, we present a private-key homomorphic encryption protocol that supports a broad class of computations, including and extending beyond BosonSampling, while providing information theoretic security guarantees. The protocol we present ensures a gap between the information accessible to an adversary and actual information encoded that grows as $m \log_2(d/m)+m(\log 2)^{-1}$ bits when $m \log_2 d$ bits are encrypted using $m$ $d$-level systems. This is a significantly stronger security guarantee than that offered by the scheme presented in \cite{PhysRevLett.109.150501}. We present our results in three parts. First we present a general approach to homomorphic encryption stemming from the group theoretic structure of quantum operations. We then present a family of operations which allow for a broad class of computations to be performed on encrypted data for a range of encryption schemes satisfying certain symmetry constraints. Finally we present a concrete encoding satisfying these constraints and show that it limits the accessible information as described above.

\textit{Group theoretic approach ---} We approach the problem of creating a homomorphic encryption scheme via the most naive route: we try to construct a set of encryption operations which commute with the operations used to implement computation on the encrypted data. However, this approach immediately encounters a barrier when applied to the case of universal computation. In such a case the computation operations form a group, either the unitary group in the case of quantum computation or the symmetric group in the case of classical reversible computation, which does not usually commute with other operations. 
Indeed, any irreducible representation of these groups only commutes with operators proportional to the identity, precluding non-trivial encryption. However, for reducible representations of these groups, there can exist non-trivial operators which commute with the entire group. 
This provides a natural route to constructing a homomorphic encryption scheme which allows the evaluation of operators chosen from some group $G$ on encrypted data, by choosing a representation of the group with a non-trivial centralizer. The set of operations used to perform the encryption must be chosen as a subset of this centralizer. While it is not immediately obvious that encryption operations chosen this way should actually be able to hide information, the BosonSampling scheme presented in \cite{PhysRevLett.109.150501} provides an example of such an encoding where a non-trivial amount of information is hidden.

\begin{figure}
        \includegraphics[width=6cm]{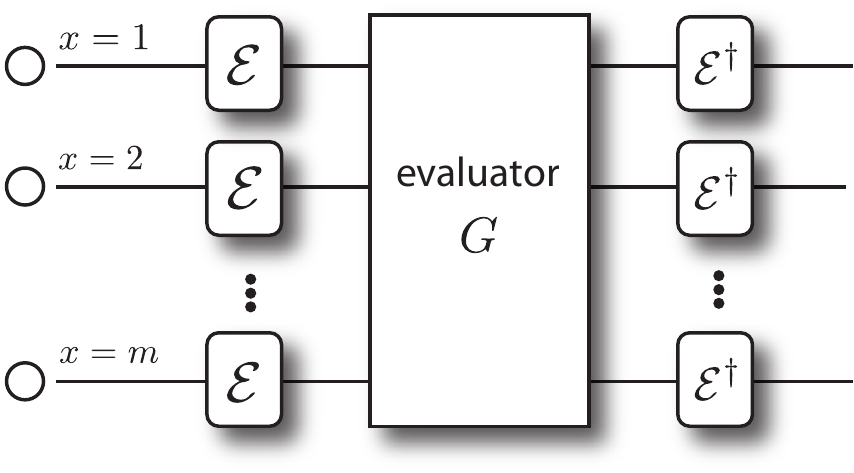}
	\caption{This figure shows Alice's encoding scheme for $m$ bosonic particles each in one of $d$ internal states. Each particle has a spatial degree of freedom labeled by $x$. The encoding operation $\mathcal{E}$ is effected across the particles in a tensor product way. 
	The evaluation operation is taken from the group $G$,
	which acts non-trivially only on the spatial modes of the $m$ bosons,
	and can put multiple bosons in a single spatial mode. 
	Post-evaluation, the encryption is removed via the inverse encoding operation to reveal the evaluated plaintext.
	}
    \label{fig1:encoding}
\end{figure}

\textit{Representation of computation ---} 
Our protocol uses $m$ identical bosonic particles; each particle has a spatial degree of freedom limited to a finite number of modes
$x =1, \dots , m $
and an internal state 
$\alpha  = 0 , \dots, d-1 $ (see Fig.~\ref{fig1:encoding}).
We design our scheme such that the encryption operations affect only the internal states of the particles, and the computation operations affect only the spatial modes of the particles. Since the input to the computation is supplied using the internal states of the particles, but the computation is performed using manipulation of only spatial modes, it may appear that the input does not affect the computation. This is not the case, however, since the internal states of the particles affect the computation by altering interference between particles.

Each particle can be represented as a state $|\alpha\>_x$ created out from a vacuum state $|\rm{vac}\>$ via a creation operator $\hat{a}_{x,\alpha}^\dag$, with $\ket{\alpha}_x=
\hat{a}_{x,\alpha}^\dag  \ket{\text{vac}}$.
The bosonic creation operators 
$\hat a _{x, \alpha} ^\dagger$ 
and 
$\hat a _{y, \beta} ^\dagger$ commute,
and satisfy the orthogonality condition \
$[a_{x,\alpha}, a_{y,\beta}^\dag]=\delta_{\alpha,\beta}\delta_{x,y}$.
Note that we make no assumption on the internal states of the $m$ particles, any two particles can have the same or different internal states.
Explicitly, the initial state of our scheme is
\begin{align}
 \hat a ^\dagger_{1, \alpha_1} \dots  \hat a ^\dagger_{1, \alpha_1} |\rm{vac}\>
 = |\alpha_1\>_1 \otimes \dots \otimes |\alpha_m\>_m, \notag
\end{align}
which we denote as $| \bsym \alpha \>$ for short, where $\bsym \alpha = (\alpha_1, \dots, \alpha_m) \in \mathbb Z_d^m$ is our plaintext.
Since the values of $\alpha_1, \dots , \alpha_m$ are selected from the integers from 0 to $d-1$, there are $d^m$ possible orthogonal input states, spanning a complex Euclidean space $  (\mathbb{C}^d)^{\otimes m}$. 

The set of computation operations that we are allowed to perform 
is isomorphic to a unitary group of a large dimension.
The state space of $m$ identical bosons can be expressed as a symmetric subspace of a Hilbert space
$\mathcal H_m = \mathcal H_{\rm internal} \otimes H_{\rm spatial}$,
where 
$\mathcal H_{\rm internal }$ 
and
$\mathcal H_{\rm spatial }$ 
denote the space for the internal degrees of freedom and 
the spatial modes of the $m$ identical bosons respectively. Due to the indistinguishability of the bosons, the state of the system is invariant under permutation of particles, and hence the system can only occupy states within the subspace of $\mathcal H_m$ which respect this permutational symmetry.
The computational operations, which act only on $\mathcal H_{ \rm spatial}$, must respect this symmetry, and hence the infinitesimal generators of the group of such operations are permutation-invariant. We proceed to elucidate the structure of these infinitesimal generators.
Each boson can be in one of $m$ possible spatial modes, and hence there are $m^2$ generalized Pauli operators each of dimension $m$ that act non-trivially on the spatial degree of freedom of each boson.
Let the corresponding Hermitian and non-Hermitian generalized Pauli operators constitute the sets $\mathcal B_i$ and $\mathcal B'_i$ respectively.
Let $\mathcal C'_i \subset \mathcal B'_i$ such that $|\mathcal C'_i|  =\frac{ |\mathcal B_i'| }{2}$
and every element in $\mathcal B'_i$ is either in or
proportional to the Hermitian conjugate of some element in $\mathcal C'_i$.
The Hermitian set 
$\overline {\mathcal B}_i 
= \mathcal B_i \cup 
	\{ P + P ^\dagger : P \in \mathcal C'_i\} \cup 
	\{ i(P - P ^\dagger) : P \in \mathcal C'_i\}$
	then comprises of $m^2$ infinitesimal generators of the unitary group operating non-trivially only on the spatial modes on the $i$-th boson.
The infinitesimal generators of group of computation operations are then symmetric sums 
of the $m$-fold tensor product of elements from $\overline {\mathcal B}_i$, 
with each such element corresponding to one boson.
The number of such symmetric sums is exactly the number of ways to 
distribute $m$ indistinguishable spatial labels (because of the requirement of permutation-invariance) among $m^2$ distinct elements of $\overline {\mathcal B}_i$,
which is $\bi{m^2 + m - 1}{m }$.
Hence the set of computation operations $G$ that we can perform is isomorphic to a unitary group of dimension at least $\bi{m^2 + m - 1}{m } \ge \frac{(m^2)^m}{m!} \ge m^m e^{m-1} / \sqrt{m}$.

Contained within $G$ are unitaries generated by the following infinitesimal generators:
\begin{align*}
\widehat{C}_{x,y}:=\sum^{d-1}_{\alpha=0}\hat{a}_{x,\alpha}^\dag \hat{a}_{y,\alpha}\ ,
\end{align*}
for $1 \leq x, y \leq m$. These operators $\widehat C_{x,y}$ are infinitesimal generators for operations that are equivalent to beam-splitters for $x \neq y$, and phase-shifters for $x=y$ in the quantum optics setting.
Since we can generate the phase-shifters and the beam-splitters as in \cite{reck1994experimental},
these infinitesimal generators generate a dimension $m$ unitary group isomorphic to U$(m)$ \cite{RevModPhys.63.375,RSG99, iachello} from which the evaluator's computation operations can be chosen. 
These are the same elements used to construct those of the BosonSampling model.
All particles in the BosonSampling model are indistinguishable (have the same internal states); 
the particles in our model however need not be indistinguishable, because each particle can be chosen as a $d$-level system independently. If we were to filter out particles with one of the $d$ internal states, we are left with a system that is equivalent to $d-1$ BosonSampling models by linearity of passive linear optics. This is a generalization of the insight used to encrypt BosonSampling instances in \cite{PhysRevLett.109.150501}.

Hence our computation space includes a hard sampling problem as a special case. However, it is currently unknown whether our model allows for encoded universal computation on a space of size exponential in $m$.

\textit{Encoding scheme ---}
For the encryption operation, a unitary operator $\mathcal{E}$, is applied to the internal state of the $m$ particles as
is depicted in Fig.~\ref{fig1:encoding}. Since $\mathcal{E}$ only acts on the internal states of the particles, {provided that it operates identically on all particles,} it commutes with our computation operations that act trivially on the internal states of the particles. In this section, we give a specific choice $\mathcal{E}$ which enables non-trivial hiding of information.

In what follows, we drop the spatial labels of the particles and make them implicit. We define the computational
 basis states of each particle to be $|\alpha\>$  for $\alpha=0,\ldots,d-1$, and define the discrete Fourier transform on $\mathbb C^d$ as
\begin{align}
F =\sum_{\alpha, \beta=0}^{d-1}\frac{1}{\sqrt{d}}\exp\left(\frac{2\pi i  \alpha \beta}{d}\right) 
|\beta\>\< \alpha|.
\notag
\end{align}
Denote the basis states of $\mathbb C^d$ in the Fourier transform basis as 
$|\alpha_F\>=F |\alpha\>$, 
and define the trigonometric terms
$c_\alpha(k) = \cos(2\pi \alpha k /d)$ and 
$s_\alpha(k) = \sin(2\pi \alpha k /d)$
for arbitrary integers $\alpha$ and $k$.
The generators of the encoding are, for $k=1,\ldots, \lfloor \frac{d}{2}\rfloor$,
\begin{align*}
\widehat{\Delta}_{k} &= 
	\frac{\widehat{L}^k+\widehat{L}^{-k}}{2}
	= 
	\sum_{\alpha=0}^{d-1 } c_\alpha (k) |\alpha_F\>\<\alpha_F|, \\ 
\widehat{\Delta}_{k+\lfloor\frac{d}{2}\rfloor} &=
-\frac{\widehat{L}^k-\widehat{L}^{-k}}{2i} 
	= 
	\sum_{\alpha=0}^{ d-1 }  s_\alpha (k)  |\alpha_F\>\<\alpha_F|,  
\end{align*}
where $\widehat{L}$ is the cyclic shift operation on the internal state of each particle such that 
$\widehat{L}\ket{\alpha}=\ket{\alpha+1({\rm mod}\ d)}$. 
To simplify our calculations, we choose to express our generators in the following basis instead: \begin{align*}
\widehat{H}_\ell
=&
\frac{1}{d}
 \left(
 	\mathbb{I} - \eta_\ell \widehat \Delta_{\lfloor \frac d 2 \rfloor} + 
 	\sum_{k = 1}^{\lfloor \frac d  2 \rfloor}
	 \left( 
	 	   2 c_\ell ( k ) \widehat \Delta_k  
	 	+ 2 s_\ell ( k ) \widehat \Delta_{k  + \lfloor \frac d 2 \rfloor} 
	 \right)
\right),
\end{align*}
where $\eta_\ell=\frac{1+(-1)^d}{2}\cos(\ell\pi)$. It is easy to verify that in the Fourier transform basis,
$\widehat H_\ell = |\ell_F\>\<\ell_F|$.

Data represented using the logical basis can be encrypted by choosing a key, $\kappa=(\kappa_1,\ldots, \kappa_{d-1})$, where each $\kappa_\ell$ is an integer chosen uniformly at random from the non-negative integers $\{0,\ldots, m\}$, and applying the random unitary operation $\mathcal E$ on each particle, where
\begin{align*}
\mathcal{E}=\exp\left(\sum_{\ell=1}^{d-1} i\phi_\ell\widehat{H}_\ell\right)\ ,
\end{align*}
and $\phi_\ell=\frac{2\pi}{m+1}\kappa_\ell$ are the secret random angles. 
It is convenient to think of $\mathcal{E}$ as a product of integer powers of $\mathcal{E}_\ell=\exp(i \widehat{H}_\ell \frac{2\pi}{m+1})$, so that $\mathcal{E}=\mathcal{E}_1^{\kappa_1}\ldots \mathcal{E}_{d-1}^{\kappa_{d-1}}$.

After the encoding, computation can still be performed on the encrypted data using the operations described in the previous section. However, for an adversary that does not have access to $\kappa$, the information encoded is obscured. Once the evaluation is completed, the output can be decrypted by applying $\mathcal{E}^\dagger$ on every particle to yield the processed plaintext. Surprisingly, with this simple encryption-decryption process, {\it any} quantum computation chosen from $G$ which is performed on the encrypted state yields the same result when decrypted, as if it were performed on the unencrypted state. The result is an encryption scheme that admits privacy homomorphisms for operations chosen from $G$.

Our scheme works because the encryption operators affect only the internal states of the particles at each site, while the computation leaves the internal states of every particle invariant. In the particular encryption scheme we have chosen, the encryption operators generate an abelian group $A$ that acts trivially on the spatial modes. Hence the evaluator can perform operations in the tensor product of the group $G$ and the abelian group $A$.

\textit{Hidden information ---} Here we show that our quantum homomorphic scheme can hide a number of bits proportional to $m$. Without knowing the key, the ensemble is $\{\hat{\rho}_{\bsym \alpha}, p_{\bsym \alpha}\}$ where $\bsym \alpha=  (\alpha_1,\alpha_2,\ldots, \alpha_m)$ denotes the plaintext, and the corresponding encrypted state is
\begin{align*}
\hat{\rho}_{\bsym \alpha}=\frac{1}{(m+1)^{d-1}}
\sum_{\kappa_1,\ldots ,\kappa_{d-1}=0}^{m}  
	\mathcal{E}^{\otimes m} 
\ket{\bsym \alpha} {\bra{\bsym \alpha}} 
(\mathcal{E}^\dag)^{\otimes m} \ .
\end{align*}
It is illuminating to look at the ensemble in the Fourier transform basis as here the encoding is diagonal. We can write $\hat{\rho}_{\bsym \alpha}$ in the form 
$\sum_{{\bsym \beta},{\bsym \beta'}\in\mathbb{Z}_d^m}
c_{\bsym \beta, \bsym \beta'} \ket{\bsym \beta} {\bra{\bsym \beta'}}$ and the non-zero coefficients are those for which the number of $\ell$'s in ${\bsym \beta}$ is equal to the number of $\ell$'s in ${\bsym \beta'}$ for all $\ell=1,\ldots, d-1$. Let $\mathcal{F}(\hat{O})$ denote $(F^\dag)^{\otimes m}\hat{O} F^{\otimes m}$. Then
\begin{align}\label{eq:part}
\mathcal{F}(\hat{\rho}_{\bsym \alpha})
=& \frac{1}{d^m}\sum_{{\bsym \beta},{\bsym \beta'}\in\mathbb{Z}_d^m}  
e^{-\frac{2\pi i {\bsym \alpha}\cdot({\bsym \beta-\bsym\beta'})}{d}} 
\ket{\bsym \beta} {\bra{\bsym\beta'}} \times
\notag\\
&\quad\prod_{\ell=0}^{d-1}\delta({\rm wt}_{\ell}({\bsym \beta})-{\rm wt}_{\ell}({\bsym \beta'}))\ ,
\end{align}
where ${\rm wt}_{\ell}({\bsym \beta})$ is the Lee weight which counts the number of times $\ell$ appears in the vector ${\bsym \beta}$. The non-zero terms in eq.~(\ref{eq:part}) can be partitioned into sets labeled by integer partitions of $m$. Let $P_{m,d}$ be the set of integer partitions of $m$ into $d$ (possibly empty) parts and let $\lambda$ be a partition in $P_{m,d}$. In eq.~(\ref{eq:part}), strings for which all Lee weights are equal belong to the same partition $\lambda$. The entries in $\lambda=(\lambda_0,\lambda_1,\ldots, \lambda_{d-1})$ give the number of times a particular element appears in ${\bsym \beta}$. With this notation, we get
\begin{align*}
\mathcal{F}(\hat{\rho}_{\bsym \alpha})
=\frac{1}{d^m}\sum_{\lambda\in P_{m,d}} R_\lambda \ket{\Psi_\lambda^{\bsym \alpha}}\bra{\Psi_\lambda^{\bsym \alpha}}\ ,
\end{align*}
where $R_\lambda=\left(\begin{array}{c} m\\ \lambda_0, \lambda_1,\ldots , \lambda_{d-1}\end{array} \right )$ is the multinomial coefficient, and
\begin{align*}
\ket{\Psi_\lambda^{\bsym \alpha}}=\frac{1}{\sqrt{R_\lambda}}\sum_{\substack{\bsym \beta: {\rm wt}_j(\bsym \beta)=\lambda_j\\ j=0,\ldots, d-1}}e^{-\frac{2\pi i }{d}{\bsym \alpha}\cdot{\bsym \beta}}
\ket{\bsym \beta}\ ,
\end{align*}
which is invariant under permutation of the particles.

{\bf Theorem 1:} For all probability distributions $p_{\bsym\alpha}$ over plaintexts $\bsym\alpha$, the accessible information of the encoding, without knowing the key, is upper bounded by $\log_2 m!$ bits when Alice sends $m$ $d$-level particles.

Proof: First, we observe that the elements of $\{\ket{\alpha}, \alpha=0,\ldots, d-1\}$ are related by powers of $\widehat{L}$. Since $\widehat{L}$ is unitary and commutes with the encoding $\mathcal{E}$, it must be that $S(\hat{\rho}_{\bsym \alpha})$ is the same for all ${\bsym \alpha}$. For simplicity, we analyze $S(\hat{\rho}_{\bf 0})$:
\begin{align}
S(\hat{\rho}_{\bf 0})&=S(\mathcal{F}(\hat{\rho}_{\bf 0}))\nonumber\\
&=
S\left(\sum_{\lambda\in P_{m,d}} \frac{ R_\lambda}{d^m} \ket{\Psi_\lambda^{{\bf 0}}}\bra{\Psi_\lambda^{{\bf 0}}}\right)
\nonumber \\
&=H\left(\left\{\frac{R_\lambda}{d^m}\right\}\right)+\sum_{\lambda\in P_{m,d}}\frac{R_\lambda}{d^m} S\left(\ket{\Psi_\lambda^{{\bf 0}}}\bra{\Psi_\lambda^{{\bf 0}}}\right)  
\nonumber\\
&= 
H\left(\left\{\frac{R_\lambda}{d^m}\right\}\right) \ ,\label{rank1}
\end{align}
where we have used the orthogonality of the different partitions labelled by $\lambda$ in third equality \cite{NielsenChuang}, and that $\ket{\Psi_\lambda^{\bf 0}}\bra{\Psi_\lambda^{\bf 0}}$ has rank one in the final equality. Similar arguments can be made for $\hat{\rho}=\sum_{\bsym \alpha}p_{\bsym \alpha}\hat{\rho}_{\bsym \alpha}$,
\begin{align}
\hspace*{-0.15cm}S(\hat{\rho})
&=
S\left(\sum_{\bsym \alpha \in \mathbb{Z}_d^m}
	p_{\bsym\alpha}
	\sum_{\lambda\in P_{m,d}}
		 \frac{R_\lambda}{d^m} \ket{\Psi_\lambda^{\bsym \alpha}}
		 	\bra{\Psi_\lambda^{\bsym \alpha}}
\right)\nonumber \\
&\le
S\left(\sum_{\bsym \alpha \in \mathbb{Z}_d^m}
	\frac 1 {d^m}
	\sum_{\lambda\in P_{m,d}}
		 \frac{R_\lambda}{d^m} \ket{\Psi_\lambda^{\bsym \alpha}}
		 	\bra{\Psi_\lambda^{\bsym \alpha}}
\right)\nonumber \\
&= H\left(\left\{ \frac{R_\lambda}{d^m}\right\}\right)
+
	\sum_{\lambda}
		 \frac{R_\lambda}{d^m} 
			S(\sum_{{\bsym\alpha}}
				\frac 1 {d^m} \ket{\Psi_\lambda^{\bsym \alpha}}
				\bra{\Psi_\lambda^{\bsym \alpha}}
			).\label{ortho2}
\end{align}
The inequality above occurs because applying a 
channel that randomizes over $\bsym \alpha$, by applying a random power of $\widehat L$ to each particle, 
symmetrizes the probability distribution $p_{\bsym \alpha}$ to the uniform distribution, but cannot decrease entropy.
The second term of eq.~(\ref{ortho2}) obeys the identity
\begin{align}
&\frac{1}{d^m}\sum_{{\bsym \alpha} \in \mathbb{Z}_d^m}\ket{\Psi_\lambda^{\bsym \alpha}}
\bra{\Psi_\lambda^{\bsym \alpha}}
=\frac{1}{R_\lambda}
\sum_{\substack{\bsym \beta: {\rm wt}_{j}({\bsym \beta})
=\lambda_j\\ j=0,\ldots, d-1}}
\ket{\bsym \beta} {\bra{\bsym \beta}}
\ ,\label{part}
\end{align} 
and is hence a maximally mixed state in the partition labeled by $\lambda$ with a rank of $R_\lambda$, with entropy at most $\max_\lambda\log_2 R_\lambda \leq \log_2 m!$.
Using these facts and putting eqs.~(\ref{rank1})-(\ref{part}) together, we obtain a bound on the Holevo quantity of
\begin{align}\label{ub}
\chi(\left \{\hat{\rho}_{\bsym \alpha},p_{\bsym \alpha}\right \}) &\leq \log_2 m! \,
\end{align}
which in turn bounds the accessible information.

When $m$ is large,
\begin{align*}
\chi(\left \{\hat{\rho}_{\bsym \alpha},p_{\bsym \alpha}\right \}) &\leq m\log_2 m -\frac{1}{\log 2} m  +\mathcal{O}(\log (m)) \ .
\end{align*}
and the gap, between the encoded information and the information accessible to an adversary, is at least
\begin{align*}
\Gamma &= m\log_2 d-\chi(\{\hat{\rho}_{\bsym \alpha},p_{\bsym \alpha}\})\\
&\approx m\log_2 (d/m)+m(\log 2)^{-1}\ .
\end{align*}
Thus if $d=m$ and $m\log_2 m$ bits are encoded, this gap scales at least proportional to $m$.
Moreover if $d = m^{1/r}$ for $r$ in the open unit interval, the gap asymptotically approaches $m(1-r)$.
  This is a significantly stronger security than that offered by \cite{PhysRevLett.109.150501}, while at the same time significantly extending the functionality by allowing computations beyond BosonSampling to be performed on the encrypted data, thus bringing us closer to the goal of achieving a quantum fully homomorphic encryption scheme. As our bound in eq.~(\ref{ub}) is independent of the probability distribution used for the encoding, the bound on the accessible information holds even if the {\it a priori} distribution on the plaintext is not uniform.

\textit{Acknowledgements ---} We thank H.~de~Guise for helpful discussions, and Y.~Li for useful comments. This material is based on research supported in part by the Singapore National Research Foundation under NRF Award No. NRF-NRFF2013-01. LC was partially supported by the Fundamental Research Funds for the Central Universities.

\bibliography{qhe_universal}
\bibliographystyle{apsrev4-1.bst}
\end{document}